\documentstyle[12pt]{article}
\newcommand{\beq}{\begin{equation}}
\newcommand{\eeq}{\end{equation}}

\def\half{{\textstyle{1\over2}}}

\def\p1half{{\textstyle{{{p+1}\over{2}}}}}

\def\23phalf{{\textstyle{{{23-p}\over{2}}}}}

\begin{document}
\thispagestyle{empty}
\begin{titlepage}

\bigskip
\hskip 3.7in{\vbox{\baselineskip12pt
%\hbox{hep-th/0105244}
}}

\bigskip\bigskip\bigskip\bigskip
\centerline{\large\bf A Proposal for Altering the Unification
Scale in String Theory}

\bigskip\bigskip
\bigskip\bigskip
\centerline{\bf Shyamoli Chaudhuri\footnote{Address after Jan 12,
2005: 1312 Oak Drive, Blacksburg, VA 24060. Email:
shyamolic@yahoo.com}} \centerline{214 North Allegheny St.}
\centerline{Bellefonte, PA 16823}
\date{\today}

\bigskip\bigskip
\begin{abstract}
An ensemble of short open strings in equilibrium with the heat
bath provided by the Euclidean worldvolume of a stack of Dbranes
undergoes a thermal phase transition to a long string phase. The
transition temperature is just below the string scale. We point
out that this phenomenon provides a simple mechanism within open
and closed string theories for altering the strong-electro-weak
coupling unification scale relative to the fundamental closed
string mass scale in spacetimes with external electromagnetic
background.
\end{abstract}
\end{titlepage}

\section{Introduction}

\vskip 0.1in There have been many attempts to explore the
viability of weakly coupled supersymmetric string theory ground
states with particle content roughly that of the minimal
supersymmetric standard model
\cite{polbook,chl,ibanez,lust,koko,kklt}. On the one hand, as
reiterated in many recent precision phenomenological analyses, the
trend in experimental data over many years overwhelmingly
continues to favour theoretical particle-cosmology models that are
close to {\em minimal} \cite{wein}. Such models traditionally
assumed a grand desert between the weak, or TeV, scale and the
coupling unification scale, which is of order $10^{16}$ GeV.
Moreover, the weak scale can be generated radiatively from the
unification scale in good agreement with precision RG analysis of
the flow of couplings in the minimal supersymmetric standard model
with three generations and a heavy top quark. On the other hand,
recent directions in braneworld phenomenology have explored the
implications of large spatial dimensions \cite{larged}, of warped
spacetime metrics \cite{rs}, of hierarchies from fluxes
\cite{kklt}, and of split supersymmetry breaking
\cite{split,koko}: a wide variety of phenomenological scenarios,
and where the string scale has been assumed to range anywhere from
the GUT scale down to the TeV scale \cite{tev}. The detailed
analyses are far from conclusive and the debate is likely to
remain controversial for some time to come. In this paper, we wish
to simply point out that there is a mechanism within open and
closed string theory that can alter the gauge coupling unification
scale significantly relative to the fundamental closed string
scale, $m_s$$=$$\alpha^{\prime -1/2}$.

\vskip 0.1in The key point is that open and closed string theories
can couple to external electromagnetic background fields
\cite{otros}. Thus, the effective string tension in a vacuum with
external electromagnetic background field is altered relative to
its value in empty space \cite{ncom}. While this phenomenon is
well-known in intersecting brane models \cite{ibanez,lust,koko},
we wish to focus on its possible significance in the broader
context of early Universe phenomenology. Consider the following
generic setting for a 4D braneworld model describing our early
Universe: a stack of intersecting Dbranes in the type IB string
with standard model gauge group and particle spectrum realized on
the worldvolume of the intersecting branes \cite{ibanez}. We will
examine such a braneworld model at finite temperature in the
Euclidean time prescription \cite{holo}. It is helpful to think of
the gravitational sector of the thermal type IB string as
providing a heat bath with which the ensemble of short open
strings on the branevolume is in thermal equilibrium. We work with
the canonical ensemble, in a fixed spatial volume $V$, and at
fixed temperature $1/\beta$.

\vskip 0.1in As shown in \cite{decon,holo}, the free open and
closed string gas undergoes a continuous phase transition of the
Kosterlitz-Thouless type at a temperature of order the string
scale. At low temperatures, we have a gas of short open strings
whose free energy scales as $T^{p+1}$. This is as expected, since
the low energy field theory isolated in the short open string
limit is a $p$$+$$1$-dimensional finite temperature gauge theory.
The order parameter for the phase transition in the low energy
finite temperature supersymmetric gauge theory is the coincidence
limit, $r$$\to$$0$, of the expectation value of a pair of timelike
Wilson loops at fixed spatial separation $r$ which, as shown in
\cite{decon} and more recently in \cite{holo}, has a
first-principles derivation from the worldsheet representation.
The leading contribution is independent of string coupling and is
given by a sum over oriented world-surfaces of cylindrical
topology with boundaries mapped to a pair of fixed curves winding
the Euclidean time direction. The leading contribution to the free
energy, on the other hand, comes from both oriented and unoriented
surfaces--- with topology of the cylinder, Mobius strip, or Klein
bottle, and it is independent of string coupling. The contribution
from the tours vanishes \cite{decon}, as shown in detail more
recently in \cite{relevant}. In \cite{decon,holo}, we calculate
the one-loop contributions to the free energy as well as to the
short distance correlation function of a pair of timelike
Wilson-Polyakov loops. The one-loop free energy turns out to
vanish identically as a consequence of tadpole cancellation for
the unphysical massless state in the Ramond-Ramond sector; a
vanishing dilaton tadpole is another remarkable consequence. The
remaining thermodynamic potentials are nonvanishing.

\vskip 0.1in It should be stressed at the outset that the
worldsheet analysis in \cite{relevant,holo} is restricted to the
thermal behavior of the canonical ensemble of open and closed
strings. All of our considerations are based on the oneloop vacuum
amplitude at finite temperature which is independent of dependence
on the string coupling. Thus, our results are expected to dovetail
neatly with any nonperturbative analyses that may follow in the
future. Although the phase transition itself lies well within the
regime of validity for our one-loop calculation, we cannot follow
the behavior of the long string phase to indefinitely high
temperatures since string loop corrections, and consequently,
gravitational interactions have not been examined thus far. Such
interactions may, for example, indicate a further phase transition
into a Schwarzschild black hole, but such phases are beyond the
realm of our worldsheet analysis. They necessarily require an
understanding of strong coupling effects. In addition, a formalism
for the {\em microcanonical} string ensemble seems necessary for a
fully convincing discussion \cite{holo}. We emphasize that our
analysis of the free energy of the string canonical ensemble in
\cite{holo} is fully compatible with the {\em analyticity} of
perturbative string amplitudes as a function of temperature; this
property is simply a special case of the well-known analytic
dependence on the moduli in any braneworld or string
compactification. On the other hand, a description of a {\em
first-order phase transition in the low energy finite temperature
supersymmetric gauge theory} can be achieved if we combine the low
energy limits of a pair of string theories linked by a thermal
T-duality transformation, namely, type IB and type I$^{\prime}$.
This phenomenon is elaborated in much more detail in our recent
papers \cite{holo,relevant}. Note that this provides an analytic
description of either side of the phase boundary at a temperature
of order the string scale.

\vskip 0.1in In recent years it has become well-known that
sub-string-scale short distance phenomena can be probed in an open
and closed string theory with D0branes--- pointlike topological
string solitons whose mass scales as $1/g$, in the presence of a
background electromagnetic field \cite{dbrane,dkps}. The key point
is the altered effective string mass scale in a vacuum with
two-form background field. Notice that the pointlike D0branes
behave like analogs of the infinitely massive heavy quarks of QCD.
Guided by this analogy, we consider parallel timelike Wilson loops
at fixed spatial separation $r$, lying in the worldvolume of
intersecting Dbranes wrapped about the Euclidean coordinate $X^0$.
The loops represent the Euclidean time world-histories of a pair
of static, semiclassical heavy quarks: the endpoints of open
string belong in the fundamental representation of the Yang-Mills
gauge group. As shown in \cite{cmnp,pairf,decon,holo}, in the
simplest case of loops lying within the D9brane stacks of the
$O(16)$$\times$$O(16)$ type IB ground state at finite temperature,
an expression for this macroscopic loop amplitude can be derived
from an extension of the usual Polyakov path integral over
connected worldsurfaces. Thus, ${\cal W}^{(2)}$ is also computed
from first principles using Riemann surface methodology, and it is
obtained by summing worldsheets with the topology of an annulus,
but with boundaries mapped to a pair of closed timelike loops,
${\cal C}_i$, ${\cal C}_f$, at fixed spatial separation $r$ in the
embedding, target-space worldvolume of Dbranes
\cite{cmnp,pairf,holo}. The result takes the form:
\begin{eqnarray}
{\cal W}_2(\beta ,r)  =&& \lim_{r \to 0} \int_0^{\infty} dt {{e^{-
r^2 t/2\pi \alpha^{\prime} }}\over{\eta(it)^{8}}}
       \sum_{n\in {\rm Z}} q^{\pi^2 n^2 \alpha^{\prime}/\beta^2
                  }
\cr
\quad && \quad \times [~ ({{\Theta_{00}(it;0)}\over{\eta(it)}})^4
 -  ({{\Theta_{10}(it;0)}\over{\eta(it)}})^4  \cr
\quad &&\quad - e^{\i \pi n}
\{ ({{\Theta_{01}(it;0)}\over{\eta(it)}})^4 -
    ({{\Theta_{11}(it;0)}\over{\eta(it)}})^4 \} ~] .
\label{eq:pairc}
\end{eqnarray}
More generally, in the generic braneworld model, the sum within
square brackets will take a more complicated form, in general
incorporating a dependence on moduli other than $\beta$
characterizing the compact spatial coordinates. But the
contribution from thermal modes will remain unchanged. The leading
$r$ dependence of the short distance pair potential is extracted
from the dimensionless amplitude ${\cal W}_2$ as follows. We set
${\cal W}_2$$=$$\lim_{\tau\to\infty} \int_{-\tau}^{+\tau} d\tau
V[r(\tau),\beta]$, inverting this relation to express $V[r,\beta]$
as an integral over the modular parameter $t$. Consider a $q$
expansion of the integrand, valid for $t$$\to$$\infty$ where the
shortest open strings dominate the modular integral. Retaining the
leading terms in the $q$ expansion and performing explicit
term-by-term integration over the worldsheet modulus, $t$,
\cite{pairf}, isolates the following short-distance pairwise
interaction \cite{decon}:
\begin{eqnarray}
V(r,\beta) =&&  (8\pi^2 \alpha^{\prime})^{-1/2} \int_0^{\infty} dt
e^{- r^2 t/2\pi \alpha^{\prime} } t^{1/2} \cr \quad &&\quad \times
\sum_{n\in {\rm Z}} ( 16 - 16 e^{i \pi n } ) q^{\alpha^{\prime}
\pi^2 n^2 /\beta^2} + \cdots \cr \quad =&& (8\pi^2
\alpha^{\prime})^{-1/2} \Gamma (3/2)
 r_{\rm min}^{3}  {{2^{5}}\over{r^3}} \left [ 1 - {{3}\over{2}} \sum_{n=0}^{\infty}
 {{ \beta_C^4 2^2 (n+\half)^2
}\over{\beta^2 r^2}} \right ] + \cdots \quad . \label{eq:static}
\end{eqnarray}
We have expressed the result in terms of the characteristic
minimum distance scale probed in the absence of external fields,
$r_{\rm min.}$$=$$2\pi \alpha^{\prime 1/2}$ \cite{dkps}, and the
bosonic closed string's self-dual inverse temperature,
$\beta_C$$=$$2\pi\alpha^{\prime 1/2}$ \cite{poltorus,polbook}. At
low temperatures, with $\beta$$>>$$\beta_C$, we can expand in a
power series, and the leading correction to the inverse power law
is $O(\beta_C^4/\beta^2 r^5)$. At high temperatures with
$\beta$$<<$$\beta_C$, the potential takes the form obtained by a
thermal duality transformation, a plausible signal indicating the
onset of the the expected long string phase. This hint is
investigated more carefully in the recent works
\cite{relevant,holo,micro}. We give evidence of a first order
deconfining phase transition in the low energy finite temperature
supersymmetric gauge theory, extracting the high temperature
behavior at temperatures far above the string scale from the low
energy limit of the Euclidean T-dual, type I$^{\prime}$ string
theory. The transition temperature is string scale.\footnote{The
precise normalization of the potential, and the transition
temperature, is given in the recent work \cite{micro}.}

\vskip 0.1in We now come to the main observation of this paper. In
the presence of an external electromagnetic field, the result
given above is modified by the simple replacements: $r_{\rm
min.}$$\to$$2\pi\alpha^{\prime 1/2}u$, $\beta_C$$\to$$u \beta_C$
where ${\cal F}^{9j}$$=$${\rm tanh}^{-1}u$ is the electromagnetic
field strength, assumed to be constant for simplicity of
calculation. The transition temperature, $T_d$$=$$T_H/u$, in the
presence of an electromagnetic background is consequently {\em
shifted} relative to the phase transition temperature in empty
space, $T_H$$=$$m_s/2{\sqrt{2}}\pi$. Since we do not as yet have
an understanding of the mechanism that might generate such an
electromagnetic field at the string mass scale, the Wilsonian
perspective suggests that we parameterize our ignorance by
interpreting this as a generic consequence of an \lq\lq effective"
string mass scale, $m_s/u$, in spacetimes with a two-form
background field \cite{ncom}. Clearly, whether one observes a
lowering, or raising, of the string mass scale depends on the
strength of the electromagnetic field. We should clarify that the
statements above only invoke the leading correction to the
effective string scale due to the background field as in
\cite{dkps}. But it is, in fact, easy to obtain the {\em exact}
electromagnetic field dependence of both the macroscopic string
amplitude, and generic one-loop string scattering amplitudes, in a
two-form background field of arbitrary strength, as has been shown
in the papers \cite{otros,ncom,pairf}.

\vskip 0.1in {\bf Note Added (Dec 2004):} Since the first
appearance of this work, we have further clarified the nature of
the deconfining thermal phase transition in the string canonical
ensemble in our papers hep-th/0409301, and hep-th/0408206. Notice
that the ensemble of short open strings in our scenario is in a
state of thermal equilibrium at the effective string scale. The
idea that a deconfining thermal phase transition might have
occurred in the early Universe, accompanied by the formation of a
{\em cosmic string} which is tentatively identified with a long
winding mode in the theory of fundamental superstrings, appears in
a 1988 paper by Englert, Orloff, and Piran
\cite{eop}.\footnote{This reference came to my notice just
recently \cite{cs2}.} An interesting point made in \cite{eop} is
the apparent clash between the necessary conditions that must be
met by cosmic string dynamics as required by galactic structure
formation vs those required by a viable beyond-the-Standard-Model
particle phenomenology. Our observation that an electromagnetic
background impacts the constraints on the latter analysis in open
and closed string theories may help in mitigating this clash,
although we do not have anything remotely close to a satisfactory
proposal for early Universe phenomenology at the moment. There has
been considerable recent activity in exploring cosmic string
dynamics in String/M theory \cite{cosmic}, although within a
different phenomenological framework that has focussed on the
important issue of moduli stabilization \cite{kklt}. We should
note that there is significant evidence for a micro-Gauss strength
magnetic field in our Universe at the super-cluster distance
scale, generally assumed by astrophysicists to be of primordial
origin \cite{freese,giov}: can this be exploited to build a viable
phenomenological model for the early Universe? The generation of
primordial magnetic fields as seeds of galaxy formation has
already played a significant role in the pre-Big-Bang Model
scenarios of Veneziano and collaborators \cite{giov}. We leave
these as tantalizing, disparate, hints that may point to a more
complete picture of the Early Universe within String/M theory in
the future.

\vskip 0.1in \noindent {\bf ACKNOWLEDGMENTS:} I thank S.\ Shenker
and M.\ Peskin for discussions in the early stages of this work,
and M.\ Giovannini, S.\ Kachru, and D.\ Lust for more recent
helpful comments. This research was funded in part by
NSF-PHY-9722394.

\end{document}